\documentclass[11pt]{article}

\usepackage{amsfonts,amsmath,amssymb,enumerate,epsfig}
\usepackage{theorem}
\usepackage{appendix}
\usepackage{color}

\numberwithin{equation}{section}
\newtheorem{definition}{Definition}[section]

\newtheorem{proposition}[definition]{Proposition}
\newtheorem{theorem}[definition]{Theorem}

\newtheorem{lemma}[definition]{Lemma}

\newcommand{\prf}{\underline{Proof:}\ }
\newcommand{\finprf}{\null \hfill {\rule{5pt}{5pt}}\\ \null}
\newcommand{\ie}{{\it i.e.}\ }
\newcommand{\lda}{\lambda}

\newcommand{\be}{\begin{equation}}
\newcommand{\ee}{\end{equation}}
\newcommand{\beu}{\begin{equation*}}
\newcommand{\eeu}{\end{equation*}}
\newcommand{\bea}{\begin{eqnarray}}
\newcommand{\eea}{\end{eqnarray}}
\newcommand{\beaa}{\begin{eqnarray*}}
\newcommand{\eeaa}{\end{eqnarray*}}
\newcommand{\bmx}{\begin{pmatrix}}
\newcommand{\emx}{\end{pmatrix}}

\def\cS{{\cal S}}         \def\cT{{\cal T}}

\newcommand{\CC}{{\mathbb C}}

\newcommand{\RR}{\mbox{${\mathbb R}$}}

\newcommand{\ZZ}{{\mathbb Z}}
\newcommand{\1}{\mbox{\hspace{.0em}1\hspace{-.24em}I}}
\newcommand{\ta}{\widetilde{a}}
\newcommand{\tb}{\widetilde{b}}
\newcommand{\trho}{\widetilde{\rho}}
\newcommand{\tA}{\widetilde{A}}
\newcommand{\tB}{\widetilde{B}}
\newcommand{\td}{\widetilde{d}}
\newcommand{\tGamma}{\widetilde{\Gamma}}
\newcommand{\tgamma}{\widetilde{\gamma}}

\advance\textheight by \topskip
\begin{document}

\renewcommand{\thefootnote}{\arabic{footnote}}
\setcounter{footnote}{0}
\newpage
\setcounter{page}{0}

\begin{center}

{\LARGE \textbf{Interplay between the Inverse Scattering Method and Fokas's Unified Transform with an Application}}

\vspace{2cm}

{\Large \textbf{Vincent Caudrelier} }

\vspace{0.5cm}

{\footnotesize School of Mathematics,
 University of Leeds,
  LS2 9JT LEEDS}

\vfill

\begin{abstract}
It is known that the initial-boundary value problem for certain integrable partial differential equations (PDEs) on the half-line with integrable boundary conditions can be mapped to a special case of the Inverse Scattering Method (ISM) on the full-line. This can 
also be established within the so-called Unified Transform (UT) of Fokas for initial-boundary value problems with linearizable boundary conditions. In this paper, we show a converse to this statement within the Ablowitz-Kaup-Newell-Segur (AKNS) scheme: the ISM on the full-line can be mapped to an initial-boundary value problem with linearizable boundary conditions. To achieve this, we need a {\it matrix} version of the UT that was introduced by the author to study integrable PDEs on star-graphs. As an application of the result, we show that the {\it new, nonlocal} reduction of the AKNS scheme introduced by Ablowitz and Musslimani to obtain the nonlocal Nonlinear Schr\"odinger (NLS) equation can be recast as an {\it old, local} reduction, thus putting the nonlocal NLS and the NLS equations on equal footing from the point of view of the reduction group theory of Mikhailov.
\end{abstract}
\end{center}
\vfill

Keywords: Inverse scattering method, unified transform, Fokas method, nonlocal NLS, reduction group.

\vfill

\noindent {\footnotesize {\tt Email: v.caudrelier@leeds.ac.uk }}\\

\newpage
\pagestyle{plain}

\section{Introduction}

Soon after the discovery of the ISM \cite{GGKM,ZS,AKNS}, the 
question of solving initial-boundary value problems for PDEs amenable to the ISM was raised. Initially, the problem on the half-line 
with specific boundary conditions at the origin was studied and connected to a special case of the ISM on the full line \cite{AS}. 
This point of view flourished over the years and took the form of a well-established theory involving a very special use of B\"acklund 
transformation as a tool to implement a nonlinear version of the method of images familiar for linear PDEs \cite{H,BT1,Fokas2,BT2,T,BH,CZ1,CZ2}. This method automatically produces so-called {\it integrable} boundary conditions \cite{Sk} on the half-line. 

If one ventures outside this well-defined arena, by considering for instance an initial-boundary value problem for an integrable PDE on the 
interval, or on the half-line with non-integrable boundary conditions or for an integrable PDE on the half-line with no natural mirror symmetry (\ie not invariant under the transformation $x\mapsto -x$, like the KdV equation), then one is led to use the so-called Unified 
Transform (UT) \cite{Fok1}. One can say that it is the appropriate generalization of the ISM to initial-boundary value problems in that it reproduces naturally all known integrable boundary conditions, called linearizable in this context. It extends naturally to problems on the interval and to PDEs with no mirror symmetry. The fundamental new insight of the method is to perform the spectral analysis of both parts of the Lax pair \cite{Lax} defining the PDE of interest.

To better understand the connection between the ISM and the UT, it was shown in \cite{BFS} how the UT with linearizable boundary conditions
is related to the mirror image approach whereby one uses the ISM on the full line with certain symmetries on the scattering data to 
obtain solutions of the PDE on the half-line which automatically satisfy the desired boundary conditions. This was done on the example 
of the NLS equation, corresponding to the $2\times 2$ AKNS scheme \cite{AKNS} with a special reduction \cite{Mikh} yielding the 
original ZS scheme for NLS \cite{ZS}. The result can be summarized as follows: the UT with linearizable boundary conditions can be mapped to a special case of the ISM on the full line\footnote{Although this was only studied explicitly for NLS, it seems reasonable that 
	the arguments of \cite{BFS} would extend to other well-known integrable PDEs within the AKNS scheme, with appropriate changes of the 
	technical details.}.

It is the objective of this article to show a converse of this statement: the ISM on the full line can be mapped to a particular case of the UT applied to an appropriate $4\times 4$ Lax pair. This provides further justification for the terminology ``unified transform'': even the ISM for problems on the full-line is just a special case of the (matrix) UT on the half-line. Note that this is 
a rather nontrivial statement since the UT is designed entirely on the half-line and knows nothing about the other half-line 
constituting the full-line. This is precisely what motivated the study in \cite{BFS} and the comparison with the mirror image method which does require the use of the ``negative'' half-line. The key is to use our matrix version of the UT. 
In fact, this matrix generalization of the UT was introduced in \cite{CCMP} in 
order to solve the open problem of formulating an ISM for integrable PDEs on star-graphs. In turn, this had been originally motivated by the difficult question of introducing local defects and impurities in classical (see e.g. \cite{BCZ,C,AD,CK,C2} and references therein) and quantum (see e.g. \cite{MRS,CMR1,CMRS,Cr} and references therein) integrable systems.
 
The main result of the paper can be worded as follows: the ISM on the full line can be mapped to a special case of the {\it matrix} UT on the half-line with linearizable boundary conditions. This is shown in Section \ref{review} after reviewing the essential ingredients of the ISM and the UT. In this paper, we work in general with the unreduced AKNS scheme and make connections with results in the reduced case producing NLS where applicable.

In Section \ref{nonlocal}, an application of our result to the nonlocal NLS equation is presented. This important byproduct establishes a precise relation between the initial value problem for the local and nonlocal NLS equation and a linearizable initial-boundary value problems for certain reductions of the matrix NLS equation. We show that our reformulation of the ISM
in terms of the UT allows one to cast the new nonlocal reduction of \cite{AM} into a standard {\it local} reduction as has been studied for several decades since \cite{Mikh}. One of the motivation for such a reformulation is that the question of classifying 
reduction groups for a given Lax pair is in general very difficult but is far more developed in the local case than in any other situation.
It led in particular to the notion of automorphic Lie algebras \cite{LM1,LM2}. In view of the enormous interest that \cite{AM} has attracted in only a few years (see e.g. \cite{AM2,AM3,F1,MSZ,GS,SG} and references therein\footnote{This is from the integrable systems point of view only but a large proportion of articles citing \cite{AM} is also related to very active area of nonlinear $PT$-symmetric models, see e.g. \cite{KYZ}.}), establishing some more systematic results about nonlocal reductions seems desirable.

\section{The matrix UT and the ISM as a special case}\label{review}

For a detailed account of the UT originally designed as the generalization of the ISM to tackle integrable PDEs on the half-line (and the interval), 
we refer the reader to the textbook \cite{Fokas}. For our purposes, we need the extension of this method designed in \cite{CCMP} to tackle integrable PDEs on a
star-graph (a collection of $N$ half-lines connected via a central vertex). There, the general $N$ case was considered in the case of the NLS reduction of the 
general AKNS scheme \cite{AKNS}. In the present paper, we will only require $N=2$ but without the NLS reduction. It turns out that all the results of \cite{CCMP} go over 
to the unreduced AKNS case with minor modifications. The proofs are completely parallel to those presented in \cite{CCMP} and are omitted in this paper. 
We collect here all the results we need with the appropriate modifications to account for the unreduced context.

We start with the following $4\times 4$ AKNS Lax pair formulation
\bea
\begin{cases}
	\label{Lax_pair1}
	\partial_x \mu+ik[\Sigma_3,\mu]=W\,\mu\,,\\
	\partial_t \mu+2ik^2[\Sigma_3,\mu]=P\,\mu\,,
\end{cases}
\eea
where 
\bea
\label{def_WP}
&&\Sigma_3=\left(\begin{array}{cc}
	\1_2 & 0\\
	0 & -\1_2
\end{array}\right)\,,~~
W(x,t)=\left(\begin{array}{cc}
	0& Q(x,t)\\
	R(x,t) & 0
\end{array}\right)\,,\\
\label{def_WP2}
&&P(x,t,k)=2k\,W-i\partial_x W\,\Sigma_3-iW^2\,\Sigma_3\,.
\eea
and 
\bea
\label{Q_diag}
Q(x,t)=\left(\begin{array}{cc}
	q^{(1)}(x,t)& 0\\
	0&q^{(2)}(x,t)
\end{array}\right)\,,~~R(x,t)=\left(\begin{array}{cc}
	r^{(1)}(x,t)& 0\\
	0&r^{(2)}(x,t)
\end{array}\right)\,.
\eea 
This gives rise to the following matrix AKNS equations
\bea
\label{matrix_AKNS_eq}
\begin{cases}
iQ_t+Q_{xx}-2QRQ=0\,,~~\\
-iR_t+R_{xx}-2RQR=0\,.
\end{cases}
\eea
The crucial differences between the ISM for problems on the line and the UT can be summarized as follows
\begin{enumerate}
	\item In the ISM, one only performs the spectral analysis of $x$-part of the Lax pair formulation at some given fixed time $t=0$ say, using the 
	initial data, while in the UT one performs the simultaneous spectral analysis of both the $x$ and $t$-part of the Lax pair formulation.
	
	\item In practice, in ISM this is achieved by introducing two fundamental (Jost) solutions normalised at $x\to\pm\infty$, giving rise to one scattering matrix $S(k)$. 
	In the UT, one uses three fundamental solutions normalised at three canonical spacetime points, giving rise to two scattering matrices $\cS(k)$, $\cT(k)$ and a 
	constraint on the scattering data known as the global relation. 
\end{enumerate}
We denote the initial-boundary data by
\bea
\label{IBV_data1}
&&Q_0(x)=Q(x,0)\,,~~R_0(x)=R(x,0)\,,~~x\ge 0\\
&&G_0(t)=Q(0,t)\,,~~G_1(t)=Q_x(0,t)\,,~~t\ge 0\\
\label{IBV_data3}
&&H_0(t)=R(0,t)\,,~~H_1(t)=R_x(0,t)\,,~~t\ge 0\,.
\eea
The data is assumed to be such that the global relation holds (see \eqref{GR1}-\eqref{GR2} below) and with appropriate smoothness conditions (e.g. $Q_0$, $R_0$ are in the Schwartz class over $\RR^+$, $G_j$, $H_j$ are smooth and with sufficient fast 
decay as $t\to\infty$). 

Note that the global relation is the crux of the matter in the UT. The method uses both the Dirichlet and Neumann data to construct the 
scattering matrix $\cT(k)$ (see \eqref{IB_data} below). However, only one of these data can be given for a well-posed problem. Therefore, one has to eliminate the other data from the general construction. The issue was addressed in \cite{CCMP} in the reduced NLS case and for $N$ half-lines (see Proposition 4.2 there). The generalization to the unreduced case simply requires to repeat the arguments twice (for $R_0$, $H_0$ and $H_1$ in addition to $Q_0$, $G_0$ and $G_1$) and the present $N=2$ unreduced case is then a particular case. Under fairly generic assumptions, the main outcome is that the Neumann data $G_1$ ($H_1$) can be expressed in terms of quantities involving only the initial condition $Q_0$ ($R_0$) and the Dirichlet data $G_0$ ($G_1$), in a way that ensures the validity of the global relation. In this article, we will not need this result as we will use certain linearizable boundary conditions which allow us to get rid of the need for the scattering data $\cT(k)$ altogether.

Define $\mu_3(x,k)$, $\mu_1(t,k)$ as the $4\times 4$ matrix-valued functions satisfying\footnote{The notations $\mu_{1,3}$ may seem a bit ad hoc here but are in line with the 
standard notations used in the analysis part of the UT under the assumption that a solution $Q(x,t)$, $R(x,t)$ exists.}
\begin{subequations}
\begin{align}
\label{def_wavefunctions1}
&\partial_x\mu_3+ik[\Sigma_3,\mu_3]=W(x,0)\,\mu_3~~,~~0<x<\infty\,,\\
\label{def_wavefunctions2}
&\partial_t\mu_1+2ik^2[\Sigma_3,\mu_1]=P(0,t,k)\,\mu_1\,,~~0<t<\infty\,,\\
\label{def_wavefunctions3}
&\displaystyle\lim_{x\to\infty}\mu_3(x,k)=\1_4\,,~~\displaystyle\lim_{t\to \infty}\mu_1(t,k)=\1_4\,.
\end{align}
\end{subequations}
The scattering matrices are defined by 
\bea
\cS(k)=\mu_3(0,k)~~\text{and}~~\cT(k)=\mu_1(0,k)\,.
\eea
 $\cS(k)$ depends on the initial data while $\cT(k)$ depends on the boundary data. They have the form
\bea
\label{IB_data}
\cS(k)=\left(\begin{array}{cc}
	\ta(k)& b(k)\\
	\tb(k)&a(k)
\end{array}\right)\,,~~\cT(k)=\left(\begin{array}{cc}
\tA(k)& B(k)\\
\tB(k)&A(k)
\end{array}\right)\,,
\eea
where all the elements shown explicitly are $2\times 2$ diagonal matrices. This yields the direct part of the method \ie the map
\be
\begin{array}{ccc}
	\{Q_0(x),R_0(x),G_0(t),G_1(t),H_0(t),H_1(t)\} & \longrightarrow & \{\cS(k),\cT(k)\}
\end{array}
\ee
The construction of the inverse map relies on the analytic properties of the scattering data as functions of $k\in\CC$ and has been shown to be most efficiently carried out by using a Riemann-Hilbert 
formulation. Specifically, define $D_j$ the $j$-th quadrant of the complex plane by
\be
D_j=\{z\in\CC,\arg z\in((j-1)\frac{\pi}{2},j\frac{\pi}{2})\}~~,~~j=1,2,3,4\,.
\ee
Then, 
\bea
&&a(k)\,,~b(k)~~\text{defined and analytic for}~\arg k\in (0,\pi)\,,\\
&&\ta(k)\,,~\tb(k)~~\text{defined and analytic for}~\arg k\in (\pi,2\pi)\,,\\
&&A(k)\,,~B(k)~~\text{defined and analytic for}~ k\in D_1\cup D_3\,,\\
&&\tA(k)\,,~\tB(k)~~\text{defined and analytic for}~ k\in D_2\cup D_4\,.
\eea
Given the scattering coefficients in $\cS(k)$ and $\cT(k)$, define the matrix $J$ by
$J(x,t,k)=J_\ell(x,t,k)$ when $\arg k=\frac{\ell\pi}{2}$, where
\bea
\label{defJ_first}
&&J_1=\left(\begin{array}{cc}
	\1_N&0\\
	\Gamma(k)\,e^{2i\phi(x,t,k)} & \1_N
\end{array}\right)\,,~~
J_4=\left(\begin{array}{cc}
	\1_N&-\gamma(k)\,e^{-2i\phi(x,t,k)}\\
	\tgamma(k)\,e^{2i\phi(x,t,k)} & \1_N-\gamma(k)\tgamma(k)
\end{array}\right)\,,\\
&&J_3=\left(\begin{array}{cc}
	\1_N&-\tGamma(k)\,e^{-2i\phi(x,t,k)}\\
	0 & \1_N
\end{array}\right)\,,~~
J_2=J_3J_4^{-1}J_1\,,
\eea
and
\bea
\label{def_gammas}
&&\gamma(k)=b(k)\,\ta^{-1}(k)\,,~~\tgamma(k)=\tb(k)\,a^{-1}(k)\,,\\
&&\Gamma(k)=\tB(k)a^{-1}(k)d^{-1}(k)\,,~~\tGamma(k)=B(k)\ta^{-1}(k)\td^{-1}(k)\,,\\
\label{defJ_last} &&d(k)=a(k)\,\tA(k)-b(k)\,\tB(k)\,,~~\td(k)=\ta(k)\,A(k)-\tb(k)\,B(k)\,.
\eea
For conciseness, in the rest of this paper, we do not consider the possibility of zeros in the scattering data as this will lengthen the paper even more with technicalities that are not essential for our results. The interested reader can refer to \cite{CCMP} for more details on such 
zeros in the star-graph case (conditions $(C1)-(C3)$ in that paper). Now, define the matrix Riemann-Hilbert problem for $M(x,t,k)$ as\footnote{Note that we do not need residue conditions here. Again, the interested reader can consult \cite{CCMP}.} 
\begin{itemize}
	\item $M$ is analytic in $k$ for $k\in\CC\setminus\{\RR\cup i\RR\}$;
	
	\item $M_-(x,t,k)=M_+(x,t,k)\,J(x,t,k)\,,~~k\in\RR\cup i\RR$
	where $M=M_-$ for $k$ in the second or fourth quadrant, $M=M_+$ for $k$ in the first or third quadrant and $J$ is defined as in 
	\eqref{defJ_first}-\eqref{defJ_last};
	
	\item $M(x,t,k)=\1_{4}+O\left(\frac{1}{k}\right)\,,~~k\to\infty$;

\end{itemize}
The form of this Riemann-Hilbert problem is dictated by an analysis of the direct part under the assumption that a solution $Q(x,t)$, $R(x,t)$ to the initial-boundary value problem exists. 
Of particular importance in this analysis is the necessary condition on the scattering data known as the global relation. In our case, it reads
\bea
\label{GR1}&&a(k)B(k)-b(k)A(k)=0\,,~~k\in D_1\,,\\
\label{GR2}&&\ta(k)\tB(k)-\tb(k)\tA(k)=0\,,~~k\in D_4\,.
\eea
We assume that the initial-boundary data is such that the global relation is satisfied. The following is a straightforward generalization 
to the present unreduced matrix case of the results of \cite{Fokas,CCMP}.
\begin{theorem}
There exists a unique solution $M(x,t,k)$ to the above Riemann-Hilbert problem. Moreover, setting
\bea
\label{recons}
W(x,t)=\begin{pmatrix}
	0 & Q(x,t)\\
	R(x,t) & 0
\end{pmatrix}=i\lim_{k\to\infty}k[\Sigma_3,M(x,t,k)]
\eea
then $Q(x,t)$, $R(x,t)$ is the solution of the matrix AKNS equations \eqref{matrix_AKNS_eq} on the half-line with initial condition $Q(x,0)=Q_0(x)$, $R(x,0)=R_0(x)$ and boundary conditions 
$Q(0,t)=G_0(t)$, $\partial_x Q(0,t)=G_1(t)$, $R(0,t)=H_0(t)$, $\partial_x R(0,t)=H_1(t)$. 
\end{theorem}

\subsection{Inverse scattering method within the UT}

In this section, we show that the standard ISM for the first nontrivial equation of the $2\times 2$ AKNS hierarchy \ie
\bea
\label{scalar_AKNS}
\begin{cases}
iq_t+q_{xx}=2qrq\,,\\
-ir_t+r_{xx}=2rqr\,,	
\end{cases}
\eea
with initial condition $q(x,0)=q_0(x)$, $r(x,0)=r_0(x)$ in the Schwartz class over $\RR$, can be seen as a special case of our matrix version of the UT with linearizable initial-boundary data (see \eqref{link_initial}-\eqref{symmetry_BC} below).
Note that this connection was already established for the reduced case of NLS ($r=\pm q^*$) in \cite{CCMP}. Eqs \eqref{scalar_AKNS} are the 
compatibility conditions of the system
\bea
\begin{cases}
\Phi_x=U\,\Phi\,,\\
\Phi_t=V\,\Phi\,,
\end{cases}
\eea
where 
\bea
&&U(x,t,k)=-ik\sigma_3+w(x,t)\,,~~V(x,t,k)=-2ik^2+p(x,t,k)\,,\\
&&p(x,t,k)=2k\,w-i\partial_x w\,\sigma_3-iw^2\,\sigma_3\,,
\eea
and
\bea
\sigma_3=\begin{pmatrix}
	1 &0\\
	0 & -1
\end{pmatrix}\,,~~w(x,t)=\begin{pmatrix}
0 & q(x,t)\\
r(x,t) & 0
\end{pmatrix}\,.
\eea 
 Equivalently, we can use $\Psi(x,t,k)=\Phi(x,t,k)e^{ikx\sigma_3}$ which satisfies
 \bea
\begin{cases}
\Psi_x+ik[\sigma_3,\Psi]=w\,\Psi\,,\\
\Psi_t+2ik^2[\sigma_3,\Psi]=p\,\Psi\,.
\end{cases} 
 \eea
 To make the connection with the usual ISM
\be
\begin{array}{ccc}
	\{q_0(x),r_0(x)\} & \longrightarrow & \{S(\lda)\}\\
	& & \downarrow\\
	\{q(x,t),r(x,t)\}& \longleftarrow& \{S(\lda,t)\}
\end{array}
\ee
we will use the Riemann-Hilbert approach to ISM instead of the historical Gelfan'd-Levitan-Marchenko formulation. It is also very convenient to consider a $4\times 4$ Lax pair formulation
\bea
\begin{cases}
	\partial_x \Psi+ik[\Sigma_3,\Psi]=W^{line}\,\Psi\,,\\
	\partial_t \Psi+2ik^2[\Sigma_3,\Psi]=P^{line}\,\Psi\,,
\end{cases}
\eea
where 
\bea
W^{line}(x,t)=\left(\begin{array}{cc}
	0& Q^{line}(x,t)\\
	R^{line}(x,t) & 0
\end{array}\right)\,.
\eea
with
 \bea
\label{def_Qline}
Q^{line}(x,t)=
\left(\begin{array}{cc}
	q(x,t) & 0\\
	0 & -q(-x,t)
\end{array}\right)\,,~~R^{line}(x,t)=
\left(\begin{array}{cc}
r(x,t) & 0\\
0 & -r(-x,t)
\end{array}\right)\,,~~x\in\RR\,.
\eea
Of course, reconstructing $q$, $r$ is equivalent to reconstructing $Q^{line}$, $R^{line}$. The advantage of using this redundant form for the standard $2\times 2$ AKNS scheme on the full line is that we will deal with
scattering data and solutions of Riemann-Hilbert problems of the same size when we compare the ISM on the full line with the matrix UT on the half-line of the previous section. 
Let us define the following fundamental solution $\Psi_+(x,k)$ of 
\bea
\partial_x \Psi+ik[\Sigma_3,\Psi]=W^{line}(x,0)\,\Psi\,,
\eea
with the normalisation
\be
\lim_{x\to\infty}\Psi_+(x,k)=\1_4\,.
\ee
The scattering data on the line is defined by
\be
\label{def_S_line}
S^{line}(k)=\lim_{x\to-\infty}e^{ikx\Sigma_3}\Psi_+(x,k)e^{-ikx\Sigma_3}\equiv \left(
\begin{array}{cc}
\ta^{line}(k) & b^{line}(k)\\
\tb^{line}(k) & a^{line}(k)
\end{array}\right)
\ee
where the four elements shown explicitly are diagonal $2\times 2$ matrices. The following result is a direct generalization to our (redundant) $4\times 4$ case of the well-known reconstruction formula 
for the ISM on the line formulated via a Riemann-Hilbert problem.
\begin{theorem}
Let $M^{line}(x,t,k)$ be the unique solution of the following Riemann-Hilbert problem\footnote{Recall that we do not consider the possibility of a discrete spectrum for conciseness.}:
\begin{enumerate}
\item $M^{line}(x,t,k)$ is an analytic function of $k$ in the upper and lower half planes, continuous from above and below the real line;

\item 
 On the real line, it satisfies a jump condition
\bea
&&M^{line}_-(x,t,k)=M^{line}_+(x,t,k)e^{-i\phi(x,t,k)\Sigma_3}J^{line}(k)e^{i\phi(x,t,k)\Sigma_3}\,,\\
&&M^{line}_\pm(x,t,k)=\lim_{\epsilon\to 0^+}M^{line}(x,t,k\pm i\epsilon)\,,~~k\in\RR\,;
\eea
where 
\bea
&&J^{line}(k)=\left(\begin{array}{cc}
\1_2 & -\rho^{line}(k)\\
\trho^{line}(k) & \1_2-\rho^{line}(k)\trho^{line}(k)
\end{array}\right)\,,\\
&&\rho^{line}(k)=b^{line}(k)(\ta^{line})^{-1}(k)\,,~~\trho^{line}(k)=\tb^{line}(k)(a^{line})^{-1}(k)\,;
\eea

\item It satisfies the normalisation condition
\be
\lim_{k\to\infty}M^{line}(x,t,k)=\1_4\,.
\ee
\end{enumerate}
Then, $Q^{line}$, $R^{line}$ defined by
\be
W^{line}(x,t)=\lim_{k\to\infty} ik[\Sigma_3,M(x,t,k)]
\ee
satisfies the matrix AKNS equations \eqref{matrix_AKNS_eq} with initial conditions 
\bea
Q^{line}_0(x)=
\left(\begin{array}{cc}
	q_0(x) & 0\\
	0 & -q_0(-x)
\end{array}\right)\,,~~R^{line}_0(x)=
\left(\begin{array}{cc}
r_0(x) & 0\\
0 & -r_0(-x)
\end{array}\right)\,.
\eea
\end{theorem}

We now show that the the problem on the full line is a special case of the matrix problem on the half-line by choosing the initial-boundary data of the latter as follows
\be
\label{link_initial}
\begin{cases}
	q^{(1)}_0(x)=q_0(x)\,,~~x\ge 0\,,\\
	q^{(2)}_0(x)=q_0(-x)\,,~~x\ge 0\,,\\
	r^{(1)}_0(x)=r_0(x)\,,~~x\ge 0\,,\\
	r^{(2)}_0(x)=r_0(-x)\,,~~x\ge 0\,.
\end{cases}
\ee
and
\be
\label{symmetry_BC}
G_0(t)-\sigma\,G_0(t)\,\sigma=0\,,~~G_1(t)+\sigma\,G_1(t)\,\sigma=0\,,
\ee
and similarly for $H_0$ and $H_1$, where
\be
\sigma=\left(\begin{array}{cc}
0&1\\
1&0	
\end{array}\right)\,.
\ee
We can now state the main result of this section
\begin{theorem}
\label{main_th}
Let $M^{red}(x,t,k)$ be the unique solution of the Riemann-Hilbert of the matrix UT described in Section \ref{review} with the particular initial-boundary data satisfying \eqref{link_initial} and \eqref{symmetry_BC}. Let 
$Q^{red}$, $R^{red}$ be the corresponding solutions of the matrix AKNS equations on the half-line defined by \eqref{recons} with $M$ replaced by $M^{red}$. 
Define $\widetilde{M}^{red}(x,t,k)$ by
\be
\widetilde{M}^{red}(x,t,k)=\begin{cases}
M^{red}(x,t,k)J_1(x,t,k)\,,~~k\in D_1\,,\\
M^{red}(x,t,k)\,,~~k\in D_2\cup D_3\,,\\
M^{red}(x,t,k)J_3^{-1}(x,t,k)\,,~~k\in D_4
\end{cases}
\ee
Then 
\be
\label{link}
\widetilde{M}^{red}(x,t,k)=I_3\,M^{line}(x,t,k)\,I_3\,,x\ge 0, k\in\CC\,.
\ee
In particular
\be
\label{link_recons}
Q^{line}(x,t)=\sigma_3Q^{red}(x,t)\,,~~R^{line}(x,t)=\sigma_3 R^{red}(x,t)\,,~~x\ge 0\,.
\ee
\end{theorem}
\prf
First note that \eqref{link_initial} and \eqref{symmetry_BC} imply the following relations on the initial-boundary scattering data
\bea
\label{relation_SSline}
&&I_3\,\cS(k)\,I_3=\Sigma I_3\,\cS(-k)\,I_3\Sigma\,S^{line}(k)\,,\\
\label{relation_T}
&& \cT(k)=\Sigma_3\,\Sigma\,\cT(-k)\,\Sigma\,\Sigma_3\,,
\eea
where $\Sigma=\1_2\otimes \sigma$ and 
\be
I_3=\left(
\begin{array}{cc}
	\sigma_3 & 0\\
	0 & \1_2
\end{array}\right)\,.
\ee
The proof is a straightforward generalisation to the present unreduced case of that of Lemma 5.1 in \cite{CCMP}. 
Next, we need to check that $\widetilde{M}^{red}$ is well-defined (same analyticity properties as $M^{red}$). We only need to 
check this in $D_1$ and $D_4$. From \eqref{defJ_first}, we see that $J_1$ only depends on $\Gamma$ which is analytic in $D_2$ by construction.
We can write 
\be
\Gamma(k)=\tB(k)\tA(k)^{-1}\frac{1}{a(k)\left(a(k)-b(k)\tB(k)\tA(k)^{-1}\right)}\,.
\ee
The global relation yields
\be
\tB(k)\tA(k)^{-1}=\tb(k)\ta(k)^{-1}
\ee
holding in $D_4$, but showing that the domain of analyticity of $\tB(k)\tA(k)^{-1}$ can be extended to $D_3$. Now \eqref{relation_T} implies 
\be
\tB(k)\tA(k)^{-1}=-\sigma \tB(-k)\tA(-k)^{-1}\sigma
\ee
which shows that $\tB(k)\tA(k)^{-1}$ can be further extended analytically to $D_1$. Therefore, $\Gamma$ is analytic in $D_1$ and hence 
$\widetilde{M}^{red}$ is well-defined and has the same analyticity properties as $M^{red}$ in that domain. The argument for $D_3$, which involves $J_3$, and hence $\tGamma$, is similar.
Now, by a direct calculation, we see that $\widetilde{M}^{red}$ is in fact continuous across $i\RR$ and only has a jump across the 
real axis given by
\be
\label{deformedRH}
\widetilde{M}^{red}_-(x,t,k)=\widetilde{M}^{red}_+(x,t,k)J_2^{-1}(x,t,k)\,,~~k\in\RR.
\ee
Also, from its definition and the asymptotic behaviour of $\cS(k)$ and $\cT(k)$ as $k\to\infty$, 
we see the normalisation of $M^{red}$ to $\1_4$ as $k\to\infty$ also holds for $\widetilde{M}^{red}$. 
Finally, we show that the jump matrix $J_2^{-1}$ coincides with $I_3\,J^{line}\,I_3$ under \eqref{relation_SSline} and \eqref{relation_T}. We have 
\be
J_2^{-1}(x,t,k)=\begin{pmatrix}
	\1_2 & -e^{-2i\phi(x,t,k)}(\gamma(k)-\tGamma(k))\\
	e^{2i\phi(x,t,k)}(\tgamma(k)-\Gamma(k))& \1_2-(\tgamma(k)-\Gamma(k))(\gamma(k)-\tGamma(k))
\end{pmatrix}
\ee
Using the global relation $\tB(k)\tA^{-1}(k)=\tb(k)\ta^{-1}(k)$ and the symmetry relation \eqref{relation_T}, we can write $\tB(k)\tA^{-1}(k)=-\sigma\tb(-k)\ta^{-1}(-k)\sigma$. From this we get
\be
\Gamma(k)=-\sigma\tb(-k)\sigma a^{-1}(k)\left[a(k)\sigma \ta(-k)\sigma+b(k)\sigma \tb(-k)\sigma \right]^{-1}\,.
\ee
Using $\tb(k)b(k)=\ta(k)a(k)-\1_2$, we derive
\be
\label{initial_dep}
\tgamma(k)-\Gamma(k)=\left[\tb(k)\sigma\ta(-k)\sigma+\ta(k)\sigma\tb(-k)\sigma  \right]\left[a(k)\sigma \ta(-k)\sigma+b(k)\sigma \tb(-k)\sigma \right]^{-1}\,.
\ee
It remains to note that \eqref{relation_SSline} yields
\bea
&&\tb^{line}(k)=\sigma_3\left[\tb(k)\sigma\ta(-k)\sigma+\ta(k)\sigma\tb(-k)\sigma  \right]\,,\\
&&a^{line}(k)=a(k)\sigma \ta(-k)\sigma+b(k)\sigma \tb(-k)\sigma\,,
\eea
to conclude that 
\be
\tgamma(k)-\Gamma(k)=\sigma_3\trho^{line}(k)\,.
\ee
Similar calculations also give $\gamma(k)-\tGamma(k)=\sigma_3\rho^{line}(k)$. Therefore, 
$\widetilde{M}^{red}$ and $I_3\,M^{line}(x,t,k)\,I_3$ satisfy exactly the same Riemann-Hilbert problem, yielding \eqref{link} by 
uniqueness of the solution. Finally, this entails \eqref{link_recons} as required since 
\be
\label{recons}
\lim_{k\to\infty} k[\Sigma_3,\widetilde{M}^{red}(x,t,k)]=\lim_{k\to\infty} k[\Sigma_3,M^{red}(x,t,k)]\,.
\ee
\finprf
Spelling out \eqref{link_recons}, we see that the problem on the full line has been entirely reconstructed as a special case of 
the matrix problem on the half-line with the following very intuitive outcome
\be
\label{map_half_full}
\begin{cases}
	q^{(1)}(x,t)=q(x,t)\,,~~x\ge 0\,,\\
	q^{(2)}(x,t)=q(-x,t)\,,~~x\ge 0\,,\\
	r^{(1)}(x,t)=r(x,t)\,,~~x\ge 0\,,\\
	r^{(2)}(x,t)=r(-x,t)\,,~~x\ge 0\,.
\end{cases}
\ee 
One may worry about the smoothness of $q$ and $r$ at $x=0$. Indeed, \eqref{symmetry_BC} only ensure that they are $C^1$ at this point. However, one can show that smoothness extends to higher orders by using the equation of motion and \eqref{symmetry_BC}.

{\bf Remarks:} 

$\bullet$ The strategy of the proof is very similar to that of \cite{BFS}. The present proof is a lot neater and more general than 
the argument presented in \cite{CCMP} in the reduced case which was a simple-minded extension of the analysis of the linear case.

$\bullet$ Put in words, the result of \cite{BFS} means that the (scalar) UT with linearizable boundary conditions can always 
be seen as a special case of the ISM on the full line with special parity conditions on the initial data. This point of view has a long 
history in the treatment of integrable PDEs with integrable boundary conditions, before the advent of the UT, as explained in the introduction.
Here, we have established a converse statement: the ISM of the full line for AKNS can be seen as a special case of the {\bf matrix} UT with linearizable boundary conditions (the conditions \eqref{symmetry_BC}).

$\bullet$ In practice, it looks like our result is not the most convenient way to approach the ISM on the full line 
as we introduce extra scattering data ($\cT(k)$) only to eliminate it in the end, using the relations \eqref{relation_SSline}-\eqref{relation_T}. However, 
at the conceptual level, our point of view is rather unifying. Firstly, it brings further justification for the use of the terminology ``unified'' transform. The central idea of a simultaneous spectral analysis of the both half of the Lax pair now also encompasses the 
historical ISM as a special case, in sharp contrast with the traditional spectral analysis of only one half of the Lax pair. Secondly, as we illustrate
in the rest of the paper on the concept of reductions, it allows us to cast ``new'' (nonlocal) reductions as ``old'' (local) ones (see below for what 
we mean by this). This produces a framework to tackle the classification of nonlocal reductions, taking advantage of the huge amount 
of available results for the local case.

$\bullet$ We should justify the terminology ``linearizable boundary conditions'' for \eqref{symmetry_BC} in our matrix problem and explain that, 
as is well-known in the scalar case, they allow us to eliminate the unknown boundary data from the reconstruction of the solution. In fact, the second point is contained in the proof of Theorem \ref{main_th}. Eq. \eqref{recons} shows that the solution for the matrix half-line problem with the conditions \eqref{symmetry_BC} can be reconstructed from the solution $\widetilde{M}^{red}(x,t,k)$ of the Riemann-Hilbert problem \eqref{deformedRH}. The latter only involves 
the initial data, through the scattering matrix $\cS(k)$, because of \eqref{initial_dep} (and the analogous relation for $\gamma(k)-\tGamma(k)$.). Now, regarding the first point, the generalization of the definition of linearizable boundary conditions to the present unreduced matrix case is as follows:  the boundary data $H_0$ and $H_1$, $G_0$ and $G_1$ is linearizable if one can find a matrix $K(k)$ such that 
\be
(-2ik^2\Sigma_3+P(0,t,-k))K(k)=K(k)(-2ik^2\Sigma_3+P(0,t,k))\,,
\ee
with $P$ given in \eqref{def_WP2}.
Let us take the simplest case where $K$ is independent of $k$. Matching the coefficients of $k$ on both side we find
\be
K=\begin{pmatrix}
	K_1 & 0\\
	0 & K_4
	\end{pmatrix}
\ee
where $K_1$ and $K_4$ are $2\times 2$ matrices, and the following boundary conditions
\be
G_0(t)K_4=-K_1G_0(t)\,,~~H_0(t)K_1=-K_4H_0(t)\,,~~G_1(t)K_4=K_1G_1(t)\,,~~H_1(t)K_1=K_4H_1(t)\,.\nonumber
\ee
The boundary conditions \eqref{symmetry_BC} correspond to choosing $K_1=\sigma=-K_4$ and are thus linearizable.

\section{The nonlocal NLS as a standard local reduction}\label{nonlocal}

\subsection{Generalities on reductions}

We need to define what we mean by ``standard'' reduction as opposed to the ``new'' reduction proposed in \cite{AM}.
The original definition of a reduction group in \cite{Mikh} can be summarized as a group $G_R$ acting on a Lax pair $U,V$ by (local) gauge 
transformations of the form, for $g\in G_R$,
\bea
&& (g\cdot U)(x,t,k)= G(x,t,k)U(x,t,\sigma_g(k))^{\#}G(x,t,k)^{-1}+\partial_xG(x,t,k)G(x,t,k)^{-1}\,,\\
&& (g\cdot V)(x,t,k)=G(x,t,k)V(x,t,\sigma_g(k))^{\#}G(x,t,k)^{-1}+\partial_tG(x,t,k)G(x,t,k)^{-1}\,,
\eea
together with an invariance requirement of the Lax pair
\be
(g\cdot U)(x,t,k)=U(x,t,k)\,,~~(g\cdot V)(x,t,k)=V(x,t,k)\,,~~\forall g\in G_R\,.
\ee
The element $G$ lives in the matrix Lie group associated to the Lie algebra of the Lax pair under consideration and the map $\sigma_g$ 
acts on the natural domain of the spectral parameter ($\CC$ for us). The operation $^{\#}$ represents possible involutions such as 
complex conjugation or transposition for instance.
The classification problem 
of reduction groups for a given Lax pair is barely tractable in general but under some assumptions, one can obtain satisfying results. 
Such considerations have deep Lie algebraic flavour and have led to the notion of automorphic Lie algebras \cite{LM1,LM2}.

We note that the gauge action is allowed to operate on the spectral parameter (via $\sigma_g$) but not on $x$ and $t$. In this sense, 
the standard reductions that have been studied for decades are local. The new, nonlocal reduction of \cite{AM}, recently vastly extended in \cite{AM2,AM3}, can be understood as a generalisation of the above gauge action where one also allows for a nontrivial action 
of the reduction group on the variables $x$ and $t$
\bea
&& (g\cdot U)(x,t,k)= G(x,t,k)U(\alpha_g(x),\beta_g(t),\sigma_g(k))^{\#}G(x,t,k)^{-1}+\partial_xG(x,t,k)G(x,t,k)^{-1}\,,\\
&& (g\cdot V)(x,t,k)=G(x,t,k)V(\alpha_g(x),\beta_g(t),\sigma_g(k))^{\#}G(x,t,k)^{-1}+\partial_tG(x,t,k)G(x,t,k)^{-1}\,,
\eea
for some maps $\alpha_g$, $\beta_g$. These maps are precisely the tools to introduce nonlocal reductions. 
For our purposes, we now assume that $G$ does not depend on $x$ and $t$. This is the most studied case of reduction groups. 
We also assume $\beta_g=id$ for all $g\in G_R$. This corresponds to the original nonlocal reduction which involved only $x$ and reads
explicitly as
\be
r(x,t)=\epsilon q^*(-x,t)\,,~~\epsilon=\pm 1\,.
\ee
When using the $2\times 2$ AKNS Lax matrix 
\be
U(x,t,k)=\begin{pmatrix}
	-ik & q(x,t)\\
	r(x,t) & ik
\end{pmatrix}
\ee
it corresponds to the following {\it nonlocal} action of the $\ZZ_2$ group with generator $s$
\be
(s\cdot U)(x,t,k)=-GU^\dagger(-x,t,k^*)G^{-1}\,,~~G=\begin{pmatrix}
	\epsilon & 0\\
	0 & -1
\end{pmatrix}
\ee
and similarly on $V(x,t,k)$. So here, we see that $\alpha_s=-id$. The well-known NLS reduction
\be
r(x,t)=\epsilon q^*(x,t)\,,~~\epsilon=\pm 1\,,
\ee
corresponds to the {\it local} $\ZZ_2$ action
\be
(s\cdot U)(x,t,k)=-GU^\dagger(x,t,k^*)G^{-1}\,,~~G=\begin{pmatrix}
	\epsilon & 0\\
	0 & -1
\end{pmatrix}
\ee
and similarly on $V(x,t,k)$, with $\alpha_s=id$.

\subsection{NLS and nonlocal NLS as local $\ZZ_2$ reductions}

We want to show that the previous two reductions appear on equal footing as two different representations of a {\bf local} $\ZZ_2$ reduction, acting on the $4 \times 4$ Lax pair of our matrix AKNS system on the half-line. 
We take the action of the $\ZZ_2$ generator $s$ on our $4\times 4$ Lax matrix $U$ to be of the form
\be
\label{action_s}
(s\cdot U)(x,t,k)=\epsilon_B BU^\dagger(x,t,\sigma_s(k))B^{-1}
\ee
with $\epsilon_B=\pm 1$, $B$ an invertible matrix (independent of $k$) and $\sigma_B$ a map on $\CC$. We present now a classification 
of such representations of $\ZZ_2$ on $U$ subject to additional requirements. The first requirement is that the 
representation be compatible with the diagonal form of our potentials $Q$ and $R$ in \eqref{Q_diag}. It turns out that this is most conveniently implemented by viewing the diagonal form \eqref{Q_diag} as resulting from a $\ZZ_2$ reduction as well. Define
the following action of the generator $\tau$ of another copy of $\ZZ_2$
\be
\label{action_t}
(\tau\cdot U)(x,t,k)=(\1\otimes \sigma_3)U(x,t,k)(\1\otimes \sigma_3)^{-1}\,.
\ee
Requiring $(\tau\cdot U)(x,t,k)=U(x,t,k)$ simply amounts to requiring $[\sigma_3,Q]=0=[\sigma_3,R]$ as desired. Now, two reductions
are compatible if their actions commute thus one requirement is that $s$ and $\tau$ commute. The other requirement comes from the 
boundary conditions \eqref{symmetry_BC}. They can also be understood as an additional $\ZZ_2$ reduction which only hold at $x=0$ and defined
by the action
\be
\label{action_u}
(u\cdot U)(0,t,k)=\Sigma U(0,t,k)\Sigma^{-1}\,,~~\Sigma=\1\otimes \sigma\,.
\ee
In particular this action should be compatible with \eqref{action_s} for $x=0$. Since the chosen representation of $s$ is independent 
of $x$, we must therefore require that $s$ and $u$ also commute in general when acting on $U(x,t,k)$. Summarizing, 
the requirements on $s$ are: it preserves the first order matrix polynomial structure of $U(x,t,k)$, 
it is an involution, it commutes with $\tau$ and it commutes with $u$.

The first condition yields $[B,\Sigma_3]=0$ and $\sigma_s(\lda)=-\frac{\lda^*}{\epsilon_B}$, so we can write
\be
B=\begin{pmatrix}
	B_+ &0\\
	0 & B_-
\end{pmatrix}
\ee
where $B_\pm$ are $2\times 2$ matrices. The involution property yields
\be
B^\dagger B^{-1}=b\1\,,~~\sigma_s^2=id\,,
\ee
for some $b\in\CC$.
The involutivity of $\sigma_s$ is automatically ensured by $|\epsilon_B|^2=1$. For consistency, we must also have $|b|^2=1$ 
so we set 
\be
b=e^{i\theta}\,,~~\theta\in\RR\,.
\ee 
Commutativity of $s$ and $\tau$ entails
\be
B(\1\otimes \sigma_3)=\gamma (\1\otimes \sigma_3)B\,,~~\gamma=\pm 1\,.
\ee
Then, commutativity of $s$ and $u$ yields
$$B\Sigma=\mu\Sigma B$$
with $|\mu|^2=1$ for consistency with $\Sigma^\dagger=\Sigma$ and $B^\dagger=bB$. The direct analysis of all these constraints yields two classes of solutions for $B$: its blocks are either diagonal or off-diagonal $2\times 2$ matrices. 
\begin{proposition}
	\label{cases_gamma}
	If $\gamma=1$ then
	\be
	B_\pm=e^{-i\frac{\theta}{2}}\rho^\pm \begin{pmatrix}
		1 & 0\\
		0 & \mu
	\end{pmatrix}\,,~~\rho^\pm\in\RR^*\,,~~\mu=\pm 1\,.
	\ee
	If $\gamma=-1$ then
\be
B_\pm=\beta^\pm \begin{pmatrix}
	0 & 1\\
	\mu & 0
\end{pmatrix}\,,~~\beta^\pm\in\RR^*~~\text{or}~~\beta^\pm\in i\RR^*\,,~~\mu=\pm 1\,.
\ee
\end{proposition}
Let us summarize our results. The action \eqref{action_s} induces the following reduction on the functions $Q(x,t)$ and $R(x,t)$
	\be
	\label{reduc}
	R(x,t)=\epsilon_B B_- Q^\dagger(x,t)B_+^{-1}\,. 
	\ee
In the case $\gamma=1$, this yields
\be
r^{(1)}(x,t)=\epsilon_B\frac{\rho^-}{\rho^+}q^{(1)*}(x,t)\,,~~r^{(2)}(x,t)=\epsilon_B\frac{\rho^-}{\rho^+}q^{(2)*}(x,t)\,,~~x\ge 0\,.
\ee 
In view of \eqref{map_half_full}, this is
\be
r(x,t)=\epsilon_B\frac{\rho^-}{\rho^+}  q^*(x,t)\,,~~x\in\RR\,.
\ee 
In the case $\gamma=-1$, we obtain
\be
r^{(1)}(x,t)=\epsilon_B\frac{\beta^-}{\beta^+}q^{(2)*}(x,t)\,,~~r^{(2)}(x,t)=\epsilon_B\frac{\beta^-}{\beta^+}q^{(1)*}(x,t)\,,~~x\ge 0\,.
\ee 
In view of \eqref{map_half_full}, this is
\be
r(x,t)=\epsilon_B\frac{\beta^-}{\beta^+}  q^*(-x,t)\,,~~x\in\RR\,.
\ee
As desired, we obtain both the local and nonlocal NLS reduction from the two allowed representations of the {\it local} reduction \eqref{action_s}.
We can now state the main result of this section.
\begin{proposition}
	The NLS equation and the nonlocal NLS equation on the full line both arise as the {\bf local} $\ZZ_2$ reduction induced by \eqref{action_s} of our matrix AKNS initial-boundary value problem \eqref{matrix_AKNS_eq} with \eqref{IBV_data1}-\eqref{IBV_data3} and \eqref{symmetry_BC} on the half-line. The NLS equation corresponds to $\gamma=1$ in Proposition \ref{cases_gamma} and the nonlocal 
	NLS equation corresponds to $\gamma=-1$.
\end{proposition}
\prf
Inserting \eqref{reduc} in \eqref{matrix_AKNS_eq} yields
\be
\label{reduced_AKNS}
iQ_t(x,t)+Q_{xx}(x,t)-2\epsilon_B (QB_- Q^\dagger B_+^{-1}Q)(x,t)=0
\ee
which holds for $x>0$ and $t>0$ as we consider an initial-boundary value problem on the half-line. By construction, the allowed reductions of Proposition \ref{cases_gamma} are compatible with the boundary conditions 
\eqref{symmetry_BC}. Therefore, Theorem \ref{main_th} applies and we can use its consequence \eqref{map_half_full}. Inserting the latter in \eqref{reduced_AKNS}, we obtain, writing out the two components explicitly:
\begin{enumerate}
\item If $\gamma=1$, for $x>0$,
\bea
&&iq_t(x,t)+q_{xx}(x,t)-2\epsilon_B\frac{\rho^-}{\rho^+}(|q|^2q)(x,t)=0\,,\\
&&iq_t(-x,t)+q_{xx}(-x,t)-2\epsilon_B\frac{\rho^-}{\rho^+}(|q|^2q)(-x,t)=0\,.
\eea
Using the smoothness of $q(x,t)$ at $x=0$, these two equations can be combined into the well-known NLS equation on the line
\be
iq_t(x,t)+q_{xx}(x,t)-2\epsilon_B\frac{\rho^-}{\rho^+}(|q|^2q)(x,t)=0\,,~~x\in\RR\,.
\ee
The real constant $\epsilon_B\frac{\rho^-}{\rho^+}$ is the usual strength of the cubic nonlinearity which can be either positive (defocusing case) or negative (focusing case). It is known that one can always rescale
$q$ to work with the canonical nonlinearity $\pm 2|q|^2q$.

\item If $\gamma=-1$, for $x>0$,
\bea
&&iq_t(x,t)+q_{xx}(x,t)-2\epsilon_B\frac{\beta^-}{\beta^+}q(x,t)q^*(-x,t)q(x,t)=0\,,\\
&&iq_t(-x,t)+q_{xx}(-x,t)-2\epsilon_B\frac{\beta^-}{\beta^+}q(-x,t)q^*(x,t)q(-x,t)=0\,.
\eea
Using the smoothness of $q(x,t)$ at $x=0$, these two equations can be combined into the nonlocal NLS equation on the line
\be
iq_t(x,t)+q_{xx}(x,t)-2\epsilon_B\frac{\beta^-}{\beta^+}q(x,t)q^*(-x,t)q(x,t)=0\,,~~x\in\RR\,.
\ee
Note that even when $\beta^\pm\in i\RR$, the ratio is real and hence we get a real coupling for the nonlocal nonlinearity. As before, it can be scaled away but its sign remains.
\end{enumerate}
\finprf
{\bf Remark:} The parameters $\theta$ and $\mu$ of $B_\pm$ play no role in the equations in both cases.

\subsection{Reduction symmetries on the scattering data}

We have established that one can obtain both the NLS and nonlocal NLS as {\it local} $\ZZ_2$ reductions applied to a matrix AKNS system on the half-line and compatible with appropriate boundary conditions. 
It is known that the NLS reduction leads to a particular symmetry of the scattering data (on the line). Similarly, it was shown in \cite{AM} that the nonlocal NLS scattering data (on the line) admits a particular symmetry which is very different from that of NLS. This is at the basis of the important differences between the solutions of these two equations. To complete our picture, we now show that these two distinct symmetries of the scattering data also emerge naturally from our reduction applied to the matrix initial-boundary value problem on the half-line. Recall that we do not consider discrete scattering data in this paper so we concentrate on the symmetries of the continuous data only. Symmetries on the discrete data are a consequence of those of the continuous data.

To derive symmetries of the scattering data on the line it is enough to consider reductions in the $x$-part of the Lax pair at some initial 
time. Compatibility with time evolution is guaranteed by construction. In our case, we need to use the reductions on both parts of the Lax pair
simultaneously. We have a priori no less than $16$ scattering coefficients (on the half-line) $a,\ta,b,\tb$, $A,\tA,B,\tB$, each of which contains $2$ coefficients. We need to show that:

\begin{itemize}
\item For $\gamma=1$ (NLS reduction), these reduce to an equivalent set of $2$ scattering coefficients on the line.
Indeed, it is well known that the NLS scattering matrix is of the generic form
\be
S_{NLS}(k)=\begin{pmatrix}
\alpha(k) & \beta(k)\\
\epsilon\beta^*(k^*) & \alpha^*(k^*)
\end{pmatrix}\,,~~\epsilon=\pm 1\,.
\ee

\item For $\gamma=-1$ (nonlocal NLS reduction), these reduce to an equivalent set of $3$ scattering coefficients on the line with additional symmetry. Indeed, in \cite{AM}, it was shown that the nonlocal NLS scattering matrix is of the generic form
\be
\label{add_sym}
S_{nNLS}(k)=\begin{pmatrix}
\alpha(k) & \beta(k)\\
\epsilon\beta^*(-k^*) & \overline{\alpha}(k)
\end{pmatrix}\,,~~\epsilon=\pm 1\,,~~\text{with}~~\alpha(k)=\alpha^*(-k^*)\,,~~\overline{\alpha}(k)=\overline{\alpha}^*(-k^*)\,.
\ee
\end{itemize}

\begin{lemma}\label{Lemma_sym}
The reduction \eqref{action_s} with $\epsilon_B=-1$ has the following consequence on the initial-boundary scattering data \eqref{IB_data}
\be
\cS^{-1}(k)=B\,\cS^\dagger(k^*)\,B^{-1}\,,~~\cT^{-1}(k)=B\,\cT^\dagger(k^*)\,B^{-1}
\ee
\end{lemma}
\prf
If $\mu$ is a solution of \eqref{Lax_pair1}, then the reduction 
\be
W(x,t)=-BW^\dagger(x,t)B^{-1}
\ee  
implies that both $\mu^{-1}(x,t,k)$ and $B\mu^\dagger(x,t,k^*)B^{-1}$ satisfies the same system
\bea
\begin{cases}
	\partial_x M+ik[\Sigma_3,M]=-M\,W\,,\\
	\partial_t M+2ik^2[\Sigma_3,M]=-M\,P\,.
\end{cases}
\eea
Applying this to $\mu_3(x,0,k)$ and using uniqueness of a normalised solution, we obtain the required symmetry for $\cS$. Similarly, 
applying this to $\mu_1(0,t,k)$ yields the required symmetry for $\cT$.
\finprf

\begin{proposition}
Under the reduction \eqref{action_s} with $B$ as in Proposition \ref{cases_gamma}, the initial-boundary data $\cS(k)$ and $\cT(k)$ in \eqref{IB_data} reduces to $2$ independent scattering coefficients for NLS and 
to $3$ scattering coefficients with the additional symmetry \eqref{add_sym} for nonlocal NLS.
\end{proposition}
\prf
We can use \eqref{relation_SSline} from the proof of Theorem \ref{main_th} and the previous lemma simultaneously.
The former  implies
\be
\label{sym_S_line1}
(S^{line})^{-1}(k)=\Sigma \,S^{line}(-k)\,\Sigma\,,
\ee
while the symmetry of $\cS$ in Lemma \ref{Lemma_sym} lifts to 
\be
\label{sym_S_line2}
(S^{line})^{-1}(k)=B\,(S^{line})^{\dagger}(k^*)\,B^{-1}\,.
\ee
Analysing the consequences of these relations on the scattering coefficients we obtain:
\begin{itemize}
	\item For $\gamma=1$:
	\be
	S^{line}(k)=\left(\begin{array}{cc|cc}
		{a^{line}}^*(k^*) & 0 & b^{line}(k) & 0\\
		0 & a^{line}(-k) & 0 & -b^{line}(-k)\\
		\hline
	-\frac{\rho^-}{\rho^+}{b^{line}}^*(k^*) & 0 & a^{line}(k) & 0\\
	0 & \frac{\rho^-}{\rho^+}{b^{line}}^*(-k^*) & 0 & {a^{line}}^*(-k^*)\\
\end{array}\right)
	\ee
	It depends only on $a^{line}(k)$ and $b^{line}(k)$ as required. From the structure of \eqref{def_Qline}, we extract from this redundant 
	$4\times 4$ matrix the required $2\times 2$ matrix  
	\be
	S_{NLS}(k)=\begin{pmatrix}
		{a^{line}}^*(k^*) & b^{line}(k)\\
		-\frac{\rho^-}{\rho^+}b^{line^*}(k^*) & a^{line}(k)
	\end{pmatrix}\,.
	\ee
The coefficient $-\frac{\rho^-}{\rho^+}$ can be rescaled to a sign as explained before.

	\item For $\gamma=-1$:
\be
S^{line}(k)=\left(\begin{array}{cc|cc}
	\ta^{line}(k) & 0 & b^{line}(k) & 0\\
	0 & a^{line}(-k) & 0 & -b^{line}(-k)\\
	\hline
	\frac{\beta^-}{\beta^+}{b^{line}}^*(-k^*) & 0 & a^{line}(k) & 0\\
	0 & -\frac{\beta^-}{\beta^+}{b^{line}}^*(k^*) & 0 & \ta^{line}(-k)
\end{array}\right)\,,
\ee
with 
\be
\label{extra}
~~a^{line}(k)={a^{line}}^*(-k^*)\,,~~\ta^{line}(k)=\ta^{line^*}(-k^*)\,.
\ee
So we have only three coefficients $a^{line}(k)$, $\ta^{line}(k)$ and $b^{line}(k)$ with the extra symmetry \eqref{extra}, as required
for nonlocal NLS. From the structure of \eqref{def_Qline}, we extract from this redundant 
$4\times 4$ matrix the required $2\times 2$ matrix  
\be
S_{nNLS}(k)=\begin{pmatrix}
	\ta^{line}(k) & b^{line}(k)\\
	\frac{\beta^-}{\beta^+}{b^{line}}^*(-k^*) & a^{line}(k)
\end{pmatrix}\,.
\ee
The coefficient $\frac{\beta^-}{\beta^+}$ can be rescaled to a sign as explained before.
\end{itemize}
\finprf

\section{Conclusion and outlook}

We have established that our matrix generalization of the UT encompasses the traditional ISM as a particular case, for the AKNS system. This is a nontrivial converse to the result obtained in \cite{BFS} (in the NLS reduction there but 
it would extend easily to the unreduced AKNS system). We have not included the case of discrete data (relevant for soliton solutions) to avoid lengthy technicalities and emphasise the important points. The discrete data could be incorporated with no essential difficulty
in all our considerations, under the usual assumptions of finite number and finite order. 
In particular, the symmetries of the discrete data follows from the symmetries of the continuous data which we discussed here in detail.

Some comments should be made in order to fully appreciate our result.
From the technical point of view, it seems like one is going through a lot of unnecessary work when using the UT with the particular symmetry \eqref{symmetry_BC} to describe the ISM of 
a problem on the full line. However, from the conceptual point of view, what this shows is that the UT is indeed a universal version of the ISM, not only as a generalisation of the latter 
for problems on the half-line or on the interval, but also as a genuine generalization of the ISM on the full line itself. We note that to see this, we had to use the $N=2$ case of the star-graph version of the UT 
first introduced in \cite{CCMP}. It is not possible to see it within the original UT which would correspond to $N=1$ in the conventions of \cite{CCMP}. 

The same universal character of our matrix UT is supported by the fact that the nonlocal reduction introduced in \cite{AM} as a new, and indeed not seen before, reduction of the standard $2\times 2$ AKNS scheme appears simply as a standard local $\ZZ_2$ reduction in our approach, provided we interpret the ISM on the line as a special case of our matrix UT on the half-line. This will allow us to establish a precise relation between initial value problem for the local and nonlocal NLS equation within a common framework of linearizable initial-boundary value problems and their connection with the mirror image method. This is left for future investigation.

An interesting question arises concerning the possibility of using the matrix UT to tackle the ``time'' and ``space-time'' nonlocal reductions introduced more recently in \cite{AM2,AM3}. We believe this should be possible as the UT naturally treats space and time parts of the 
Lax pair on equal footing. This is left for future work.

The fact that one can treat local and nonlocal reductions on equal footing, using appropriate higher dimensional matrix versions of 
traditional local reductions opens the way to a more systematic study of nonlocal reductions for various kind of multicomponent generalizations of the AKNS hierarchy, using the vast amount of knowledge accumulated on local reductions for these systems (see e.g. 
\cite{FK,GGK}). In principle, 
one ``only'' has to apply the strategy of the present paper to these more general situations: pick a reduction group, double 
the size of the representations and retain only those compatible with appropriate extra reductions analogous to \eqref{action_t} and 
\eqref{action_u}. Recent results like those in \cite{Gerd,Gur} should emerge as particular cases of this approach. A related 
open direction would be the adaptation of the present ideas to other members of the AKNS hierarchy. For instance, it is known 
that the next equation in the hierarchy yields the modified KdV equation under a standard $\ZZ_2$ reduction. It would be interesting to 
see how the nonlocal reductions for the mKdV equation introduced in \cite{AM2} fit into a local reformulation for a matrix problem on the half-line, if at all possible.

\paragraph{Acknowledgments.} It is a pleasure to acknowledge discussions with A. Mikhailov on the reduction group.

\end{document}